\documentclass[aps,prl,twocolumn,10pt,showpacs]{revtex4-1}

\usepackage[utf8x]{inputenc}
\usepackage{graphicx}
\usepackage{amsmath,amssymb}
\usepackage{datetime}
\usepackage{color}
\usepackage[svgnames]{xcolor}
\usepackage[
pdftitle={Creating State-Dependent Lattices for Ultracold Fermions by Magnetic Gradient Modulation},    
pdfauthor={Gregor Jotzu, Michael Messer, Frederik Goerg, Remi Desbuquois, Daniel Greif, Tilman Esslinger},     
pdfsubject={Optical Lattice},   
pdfcreator={Gregor Jotzu, Michael Messer, Frederik Goerg, Remi Desbuquois, Daniel Greif, Tilman Esslinger},   
pdfkeywords={Optical Lattice } {Hubbard Model } {Ultracold Fermions }, 
colorlinks=True,linkcolor=DarkSlateBlue,citecolor=DarkBlue,urlcolor=DarkBlue,
	pdfstartview=FitH,bookmarks=False,pdfpagemode=UseNone
]{hyperref}

\begin{document}

\title{Creating State-Dependent Lattices for Ultracold Fermions 
by~Magnetic~Gradient~Modulation}

\author{Gregor~Jotzu}
\author{Michael~Messer}
\author{Frederik~Görg}
\author{Daniel~Greif}
\author{Rémi~Desbuquois}
\author{Tilman~Esslinger}
\affiliation{Institute for Quantum Electronics, ETH Zurich, 8093 Zurich, Switzerland}

\begin{abstract}
We demonstrate a versatile method to create state-dependent optical lattices by applying a magnetic field gradient modulated in time.
This allows for tuning the relative amplitude and sign of the tunnelling for different internal states.
We observe substantially different momentum distributions depending on the spin-state of fermionic $^{40}\mathrm{K}$ atoms.
Using dipole-oscillations we probe the spin-dependent band structure and find good agreement with theory.
\textit{In-situ} expansion-dynamics demonstrate that one state can be completely localized whilst others remain itinerant.
A systematic study shows negligible heating and lifetimes of several seconds in the Hubbard regime.
\end{abstract}


\maketitle

Ultracold atoms in optical lattices provide a highly tunable platform to simulate the behaviour of electrons in solids. 
When the tunnelling in the lattice depends on the internal spin state, SU(2) symmetry is explicitly broken and novel quantum phases emerge, such as unconventional superconductivity owing to a Fermi-surface mismatch, or exotic forms of magnetism arising from anisotropic spin exchange\,\cite{Casalbuoni2004,Auerbach1994}.
Both the static properties and the dynamics of particles in the lattice will then depend on their internal state.
Realizing such a state-dependent tunnelling in an optical lattice requires a coupling between internal and external degrees of freedom and has so far been demonstrated with bosonic atoms, allowing for quantum computation and simulation\,\cite{Deutsch1998,Jaksch1998,Haycock2000,Mandel2003,Lee2007,Soltan-Panahi2010,Catani2009,Lamporesi2010,McKay2010,Gadway2010,Gadway2011}. 
The technique used there mainly relies on the differential coupling of the lattice laser field to different atomic transition lines 
\footnote{A mass imbalance as in Refs.\,\cite{Catani2009, Lamporesi2010} also contributes to the state-dependent tunnelling.}. 
Its range of applicability is however limited by the intrinsic problem of heating by spontaneous emission. This problem is particularly severe for fermionic potassium and lithium atoms because of their small fine-structure splittings, which has hindered the realisation of state-dependent optical lattices for fermions. 
Other proposed methods, relying on earth-alkaline and similar atoms\,\cite{Yi2008,Daley2011}, or on atom chips\,\cite{Fortagh2007}, seem to involve difficult experimental challenges.

\begin{figure}
    \includegraphics{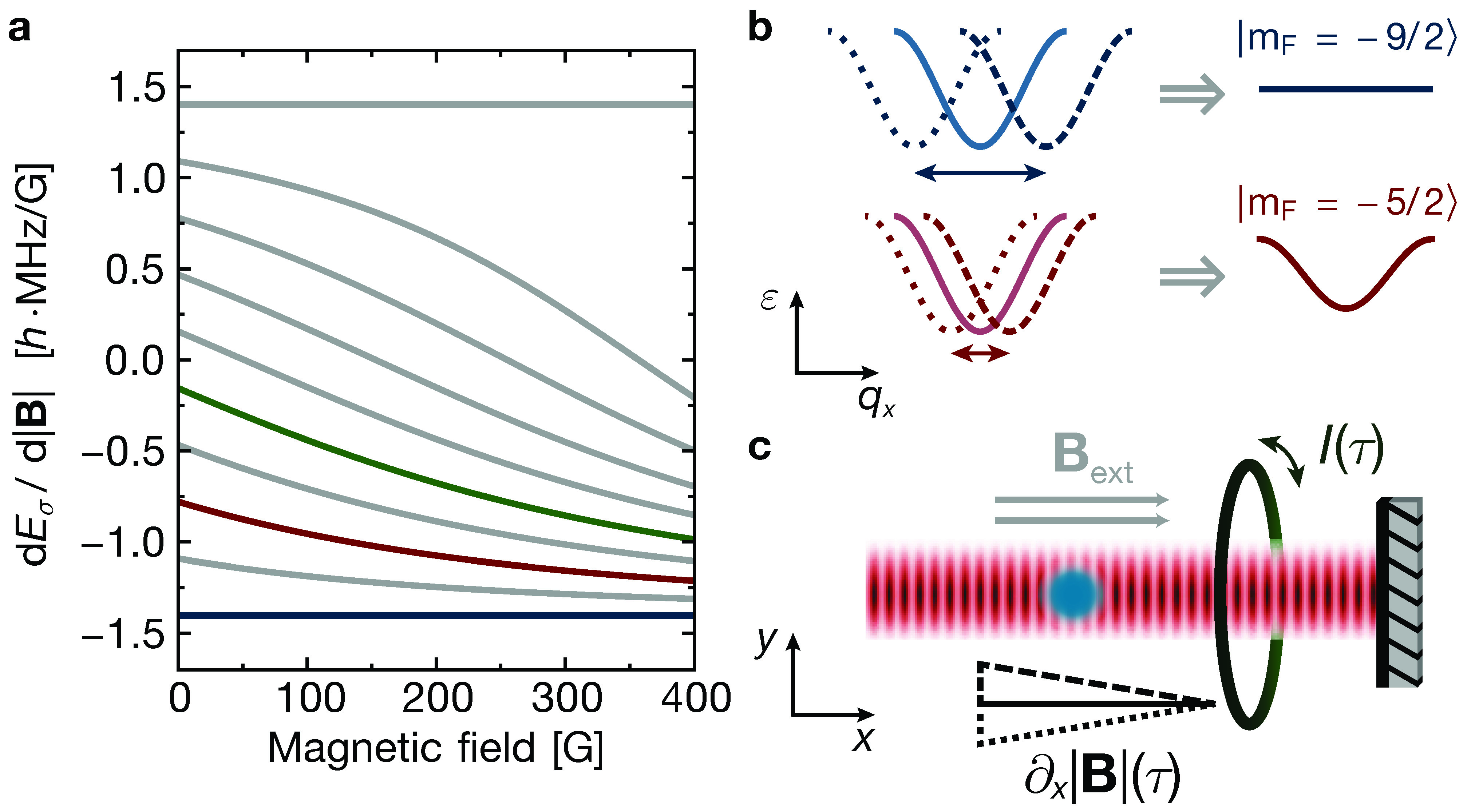}
    \caption{
    {\bf (a)} 
    Magnetic moments of the $F=9/2$ hyperfine manifold of $^{40}\mathrm{K}$ as a function of the external field. Blue, red and green denote the $\left|m_F= -9/2\right\rangle$, $\left| -5/2\right\rangle$ and $\left| -1/2\right\rangle$ sub-levels throughout.
    {\bf (b)} 
    The effective energy bands in quasimomentum space are given by the time-average of the ``shaken'' bands. As the force depends on the internal state, so does the effective band.
    {\bf (c)}
    A cloud of $^{40}\mathrm{K}$ (blue) is trapped in a retro-reflected laser beam. An oscillating current $I(\tau)$ in a single coil creates the oscillating gradient $\partial_x|\mathbf{B}|$. The uniform external field $\mathbf{B}_{\mathrm{ext}}$ is provided by additional coils.
	}\label{fig:Setup}	
\end{figure}

Here we present the implementation of a spin-dependent lattice for ultracold fermions using a different method. 
Following the proposal in Ref.\,\cite{Jotzu2014}, our method relies on the application of an oscillating force to the particles in the lattice, with an amplitude which depends on their internal state
\footnote{To create spin-orbit coupling, similar schemes were proposed\,\cite{Xu2013, Anderson2013, Struck2014} and very recently demonstrated with bosons in a harmonic trap\,\cite{Luo2015}.}.
The resulting system can then be well described by an effective time-independent Hamiltonian\,\cite{Dunlap1986,Goldman2014,Bukov2014,Eckardt2015}, with differently renormalized tunnelling terms for each internal state.
The general idea can readily be extended to mixtures of different atomic species or other artificial or conventional lattices. 
In our case the spin-dependent force  in a particular direction $x$ on state $\sigma$ is provided by a magnetic field gradient, and is given by
\begin{equation} 
F_{x,\sigma} = -\dfrac{\mathrm{d}E_{\sigma}}{\mathrm{d}|\mathbf{B}|} \partial_x|\mathbf{B}|\label{eq:SpinForce}
\end{equation}
where $E_{\sigma}$ gives the energy of a state as a function of the magnitude of the external magnetic field $|\mathbf{B}|$ 
\footnote{Throughout this work we operate in a regime where the Larmor-frequency is sufficiently high that the spin is always aligned with the external field.}.
For the $F = 9/2$ hyperfine-manifold of $^{40}\mathrm{K}$ used in our experiment, the force resulting from a given gradient can take on various positive or negative values or vanish, depending on the Zeeman-sublevel, see Fig.\,\ref{fig:Setup}a.

An oscillating force renormalizes the amplitude and phase of tunnel-couplings in a lattice; this effect has been observed in optical lattices for ultracold bosons\,\cite{Dunlap1986,Madison1998,Lignier2007,Kierig2008,Struck2012}.
In a single-band tight-binding model, the time-dependent Hamiltonian of a one-dimensional non-interacting system is given by:
\begin{equation} 
\hat{H}(\tau) = 
-t \sum_{j,\sigma} \hat{c}_{j,\sigma}^{\dagger}\hat{c}_{j+1,\sigma} 
+ \mathrm{H.c.}  
- a \sum_{\sigma} F_{x,\sigma}(\tau) \sum_{j} j\hat{c}_{j,\sigma}^{\dagger}\hat{c}_{j,\sigma} \nonumber
\label{eq:TimeDepHamiltonian}
\end{equation}
where $\hat{c}_{j,\sigma}^{\dagger}$ and $\hat{c}_{j,\sigma}$ are the creation and  annihilation operators of one fermion with spin $\sigma$ on site $j$, $\tau$ is time, $a$ the lattice constant and $t$ the tunnelling energy.
Using Floquet theory, an effective time-independent Hamiltonian can be derived, which describes the system on time-scales longer than the oscillation period\,\cite{Shirley1965,Rahav2003,Eckardt2005,Goldman2014,Bukov2014}.
For a sinusoidally oscillating force $F_{x,\sigma}(\tau) = \kappa_{\sigma}h\nu_{\mathrm{S}}/a \cdot \sin{\left(2\pi\,\nu_{\mathrm{S}}\tau\right)}$, with frequency $\nu_{\mathrm{S}}$ and a dimensionless modulation amplitude $\kappa_{\sigma}$, the effective Hamiltonian reads (see\,\cite{Dunlap1986,Goldman2014,Bukov2014,Eckardt2015} for a derivation):

\begin{equation} 
\hat{H}^{\mathrm{eff}} = 
-t\sum_{\sigma}\mathcal{J}_0(\kappa_{\sigma}) \sum_{j}\hat{c}_{j,\sigma}^{\dagger}\hat{c}_{j+1,\sigma} 
+ \mathrm{H.c.}  
\label{eq:Heff}
\end{equation}
The tunnelling has therefore been renormalized to a spin-dependent value 
\begin{equation} 
t^{\mathrm{eff}}_{\sigma} = t \mathcal{J}_0(\kappa_{\sigma})
\label{eq:teff}
\end{equation}
given by a $0^\mathrm{th}$-order ordinary Bessel function $\mathcal{J}_0$.
An intuitive picture of this effect can be gained by considering the time-dependent band energy as a function of quasimomentum $q_x$ in a co-moving frame.
As illustated in Fig.\,\ref{fig:Setup}b, the average of the ``shaken''  band energy over one period then gives the effective band:
\begin{equation} 
\epsilon_{\sigma}^{\mathrm{eff}}\left(q_x\right)  = \left\langle -2t \cos{\left( aq_x-a\int_0^{\tau} F_{x,\sigma}(\tau)\mathrm{d}\tau\right)}\right\rangle_{\tau}  
\label{eq:TimeDepBand}
\end{equation}
Going beyond the non-interacting single-band regime, these effective Hamiltonians can contain additional terms such as longer-range tunnelling, as recently observed\,\cite{Parker2013,Rechtsman2013,Jotzu2014}, which could then also be made state-dependent.

\begin{figure}
    \includegraphics{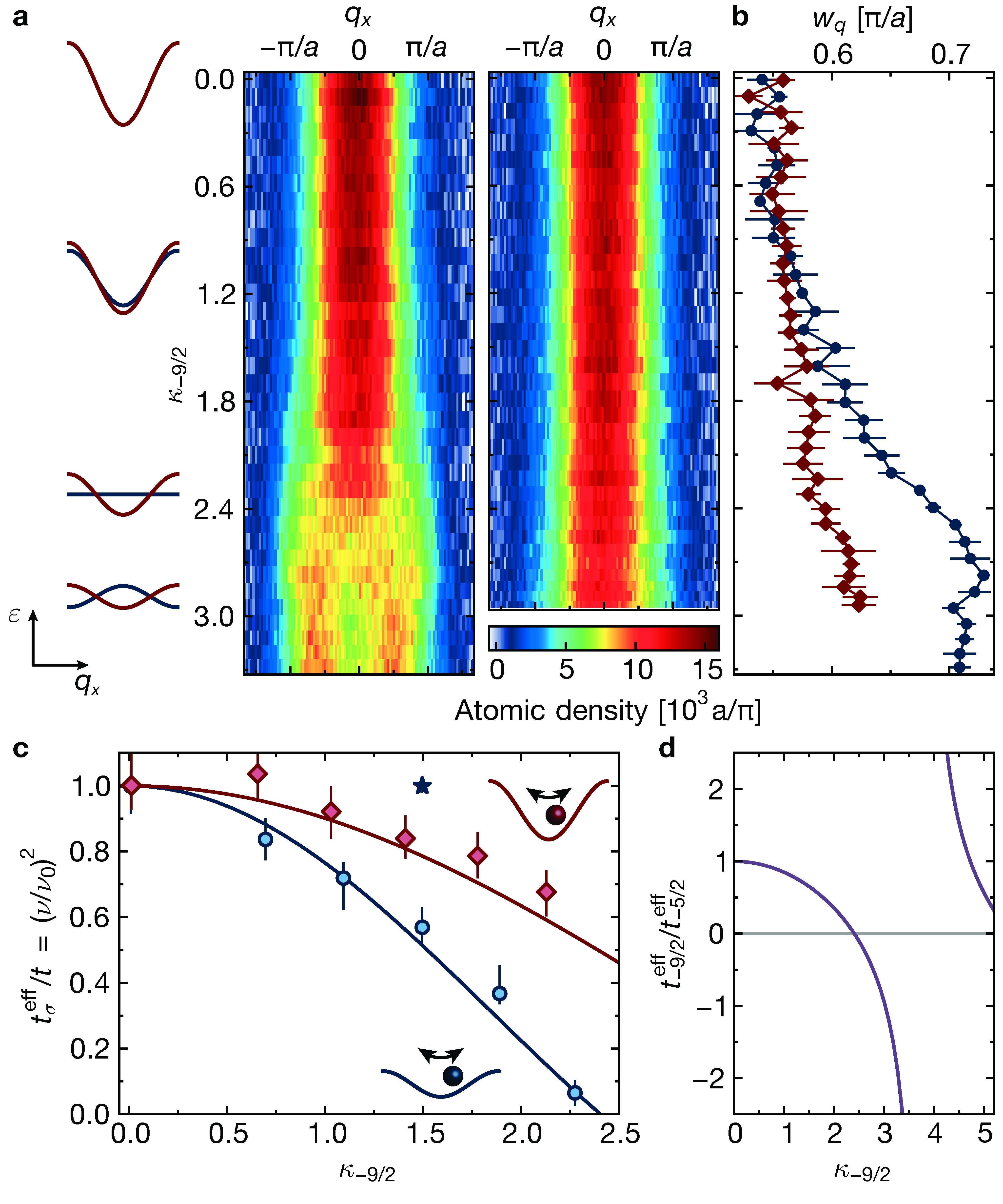}
    \caption{Fermions in spin-dependent bands.
    {\bf (a)}
	Quasimomentum distribution in the lattice, summed over $q_y$ and $q_z$, as a function of gradient modulation amplitude for the $\left| -9/2\right\rangle$ (left) and $\left| -5/2\right\rangle$ (right) state. 
	The effective band structures for each spin state are shown as diagrams for $\kappa_{-9/2}=0$, 1.2, 2.4 and~3.0.  
	{\bf (b)}
	Second moment of the distributions\,\cite{Supplemental}. 
	Blue circles (red diamonds) denote $\left| -9/2\right\rangle$ ($\left| -5/2\right\rangle$ atoms).
	Data show mean $\pm$ s.d. of 5 measurements.
	{\bf (c)}
	Dipole oscillations in the spin-dependent bands (illustrated in the inset) show how the tunnelling is renormalized for each spin, relative to the oscillations without modulation (which have frequency $\nu_0$). 
	The oscillation frequency squared is proportional to the tunnelling and is determined from fits to at least 60 measurements of the time-dependent quasimomentum peak position.
	Error bars show the fit uncertainty and solid lines the Bessel-functions for each spin, calculated without free parameters.
	The star indicates $\left| -9/2\right\rangle$ atoms without a lattice.
	{\bf (d)}
	Tunnelling ratio of the two spins as a function of modulation amplitude.
	}\label{fig:LinesumsDO}
\end{figure}

In a first experiment we study how the changing band structure of the effective Hamiltonian affects the quasimomentum distribution of two different spin states.
For this measurement, we prepare a degenerate cloud of $1.9(4)\times10^4$ non-interacting spin-polarized fermionic $^{40}\mathrm{K}$ atoms in an optical dipole trap operating at $826\,$nm and load them into the lowest band of a one-dimensional optical lattice with a lattice constant of $a = 532\,$nm and a tunnelling energy of $t~=~h\times 174(9)\,$Hz.
For details of the preparation procedure and trapping parameters, see\,\cite{Supplemental}.
We then ramp up an oscillating current with a frequency of $\nu_{\mathrm{S}}=750\,$Hz within 100ms, which runs through a single coil mounted about $1\,$cm away from the atoms, see Fig.\,\ref{fig:Setup}c. 
Reaching the first zero of $t^{\mathrm{eff}}_{-9/2}$ at $\kappa_{-9/2} \approx 2.4$ requires a gradient amplitude of about $24\,\mathrm{G}/\mathrm{cm}$ corresponding to a current amplitude of $6.4\,$A.

Fig.\,\ref{fig:LinesumsDO}a shows the resulting quasimomentum distributions in the lattice, measured using a band-mapping technique\,\cite{Supplemental}.
The distribution broadens when the modulation is increased, because the width of the lowest band decreases. 
Above the critical value of $\kappa_{-9/2}\approx2.4$, a double-peak feature appears for atoms in the $m_F = -9/2$ Zeeman sub-level (henceforth denoted $\left| -9/2\right\rangle$, and similarly for other $m_F$ values). 
This occurs because $t^{\mathrm{eff}}_{-9/2}$ becomes negative and therefore the band has minima at quasimomenta $q_x = \pm \pi/a$ rather than at $q_x=0$.
The situation is very different when using the $\left| -5/2\right\rangle$ state.
At the offset-field of $57.53(1)\,$G used here, $\kappa_{-5/2}=0.636(2)\times \kappa_{-9/2}$ for the same magnetic field gradient amplitudes.
Therefore, when the $\left| -9/2\right\rangle$ atoms experience a completely flat band, those in the $\left| -5/2\right\rangle$ state still tunnel with $t^{\mathrm{eff}}_{-5/2}=h\times 86(4)\,$Hz.
As can be seen in Fig.\,\ref{fig:LinesumsDO}b, their spread increases only slightly.
The two states experience very different effective band structures, which allows for creating a tunable Fermi-surface mismatch.

In order to directly measure the effective tunnelling, we perform dipole oscillations in the modulated optical lattice.  
The oscillation frequency $\nu$ is given by
\begin{equation} 
\nu^2 = \alpha(q_0)\frac{\pi^2}{E_{\mathrm{R}}} \nu^2_x\,t
\label{eq:DOfreq}
\end{equation}
where $\nu_x$ is the trap frequency and $E_{\mathrm{R}}=h^2/(8ma^2)$ denotes the recoil energy of the atoms with mass $m$ in the lattice\,\cite{Cataliotti2001}. The parameter $\alpha(q_0)$ describes the effect of the anharmonicity of the dispersion and depends on the initial displacement $q_0$\,\cite{Supplemental}.
The atoms are displaced by $q_0=0.31(4)\, \pi/a$ in $q_x$-direction and allowed to evolve for up to $350\,$ms.
The oscillation frequency is then extracted from the time-dependent peak position of the quasimomentum-distribution. 
The maximum oscillation frequency of $\nu=8.4(3)\,$Hz is much smaller than the modulation frequency of $750\,$Hz, meaning that the dynamics should be well described by $\hat{H}^{\mathrm{eff}}$.
As shown in Fig. \ref{fig:LinesumsDO}c, the oscillations for $\left| -9/2\right\rangle$ atoms become slower when the modulation amplitude is increased, as expected from the Bessel functions in Eq.\,\ref{eq:teff}.
The $\left| -5/2\right\rangle$ atoms, on the other hand, experience a weaker modulation force and their oscillation frequency therefore changes much less.
The spin-dependent oscillation frequency shows that the atoms behave as though they had different masses in the lattice.
The effect is expected to vanish in the absence of a lattice, as a quadratic dispersion is not changed by an oscillating force \cite{Grozdanov1988}. 
Indeed, Fig.\,\ref{fig:LinesumsDO}c shows that we observe no reduction of the oscillation frequency when applying an oscillating gradient in a harmonic trap.

\begin{figure}
    \includegraphics{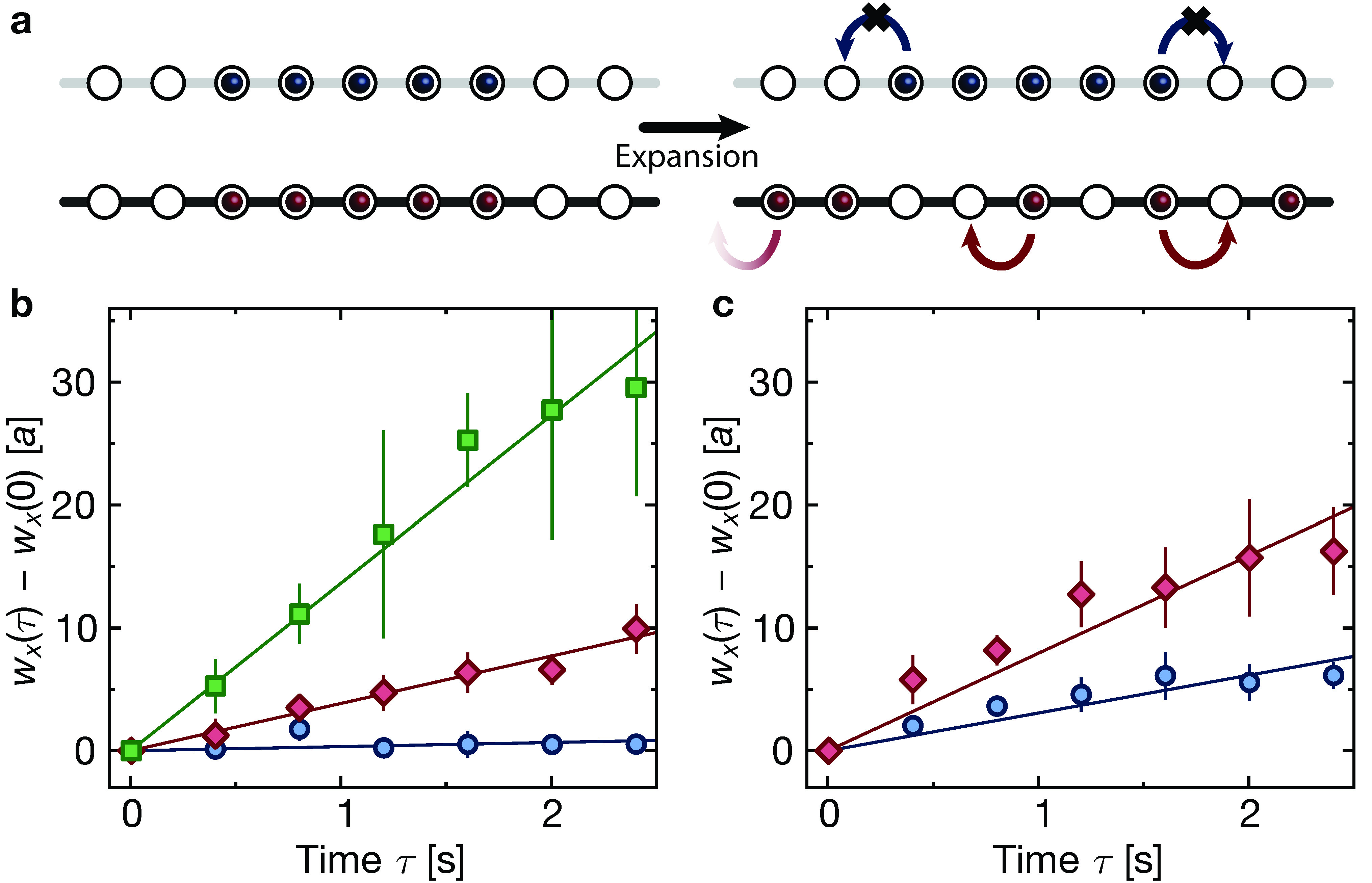}
    \caption{
    Spin-dependent expansion dynamics for $\kappa_{-9/2} = 2.41(4)$ 
    {\bf (a)} 
    Whilst the effective tunneling for the $\left| -9/2\right\rangle$ atoms is suppressed, atoms in other states are still itinerant and the cloud expands. 
    {\bf (b)} 
    Gaussian width $w_x$ of the real-space density distribution\,\cite{Supplemental} of spin-polarized non-interacting $\left| -9/2\right\rangle$ (blue circles), $\left| -5/2\right\rangle$ (red diamonds) and $\left| -1/2\right\rangle$ (green squares) atoms, compared to their initial values.
    {\bf (c)} 
    Expansion of a repulsively interacting mixture of $\left| -9/2\right\rangle$ and $\left| -5/2\right\rangle$ atoms.
    Data show mean $\pm$ s.d. of 5\,({\bf b}) or 9\,({\bf c}) measurements.
    Solid lines are linear fits.
	}\label{fig:Expansion}
\end{figure}

By tuning the modulation amplitude, the tunnelling ratio of the two spins can be set to any positive or negative value, as shown in Fig.\,\ref{fig:LinesumsDO}d.
We now focus on the case where $\kappa_{-9/2} \approx 2.4$ and the $\left| -9/2\right\rangle$ atoms experience a completely flat band with zero tunneling. 
They are therefore pinned to the lattice, whilst atoms in other states remain itinerant, see Fig.\,\ref{fig:Expansion}a.
This situation can be observed in the \textit{in-situ} expansion of the atomic cloud.
For these measurements we work with single spin states at a uniform offset-field of $208.15(1)\,$G, as typically used for experiments with interacting $^{40}\mathrm{K}$.
Starting from a harmonic trap with frequencies $\omega_{x,y,z}/2\pi=(67.8(3),60.4(4),233.5(3))\,$Hz we suddenly switch off the confinement along the lattice direction and measure the width of the cloud as a function of time\,\cite{Creffield2010,Schneider2012}.
Fig.\,\ref{fig:Expansion}b shows that no expansion of the cloud can be observed for the $\left| -9/2\right\rangle$ state, where $\kappa_{-9/2} = 2.41(4)$ and $t^{\mathrm{eff}}_{-9/2}=h\times 0(4)\,$Hz.  
However, a broadening is clearly seen for the $\left| -5/2\right\rangle$ state (where $\kappa_{-5/2} = 1.86(3)$ and $t^{\mathrm{eff}}_{-5/2}=h\times 53(3)\,$Hz) as well as the $\left| -1/2\right\rangle$ state (where $\kappa_{-1/2} = 1.19(5)$ and $t^{\mathrm{eff}}_{-1/2}=h\times 117(6)\,$Hz). 
This demonstrates spin-selective pinning, a crucial ingredient for the Falikov-Kimball model, the Kondo (lattice) model and other models for impurities or disordered systems\,\cite{Falicov1969,Kondo1964}.

We now study the expansion of an interacting spin-mixture, by simultaneously loading atoms in the $\left| -9/2\right\rangle$ and $\left| -5/2\right\rangle$ state. 
During the modulation, the scattering length of $257(1)a_0$, where $a_0$ denotes the Bohr radius, also varies between ${1280(80)}a_0$ and ${217.0(1)}a_0$
\footnote{As we create the magnetic field gradient with a single coil, we additionally modulate the magnetic field itself. Due to the vicinity of a Feshbach resonance, this leads to a large modulation of the scattering length \cite{Supplemental}}.
In Fig.\,\ref{fig:Expansion}c we see that the expansion is still very different for the two states, however the $\left| -9/2\right\rangle$ component now shows a slight broadening.
This may be caused by additional terms in $\hat{H}^\mathrm{eff}$ which have been predicted to arise owing to the presence or modulation of interactions, in particular for low frequencies\,\cite{Creffield2006,Verdeny2013,Bukov2014,Eckardt2015,Gong2009,DiLiberto2014,Greschner2014}, and is an interesting subject for further studies.

\begin{figure}
    \includegraphics{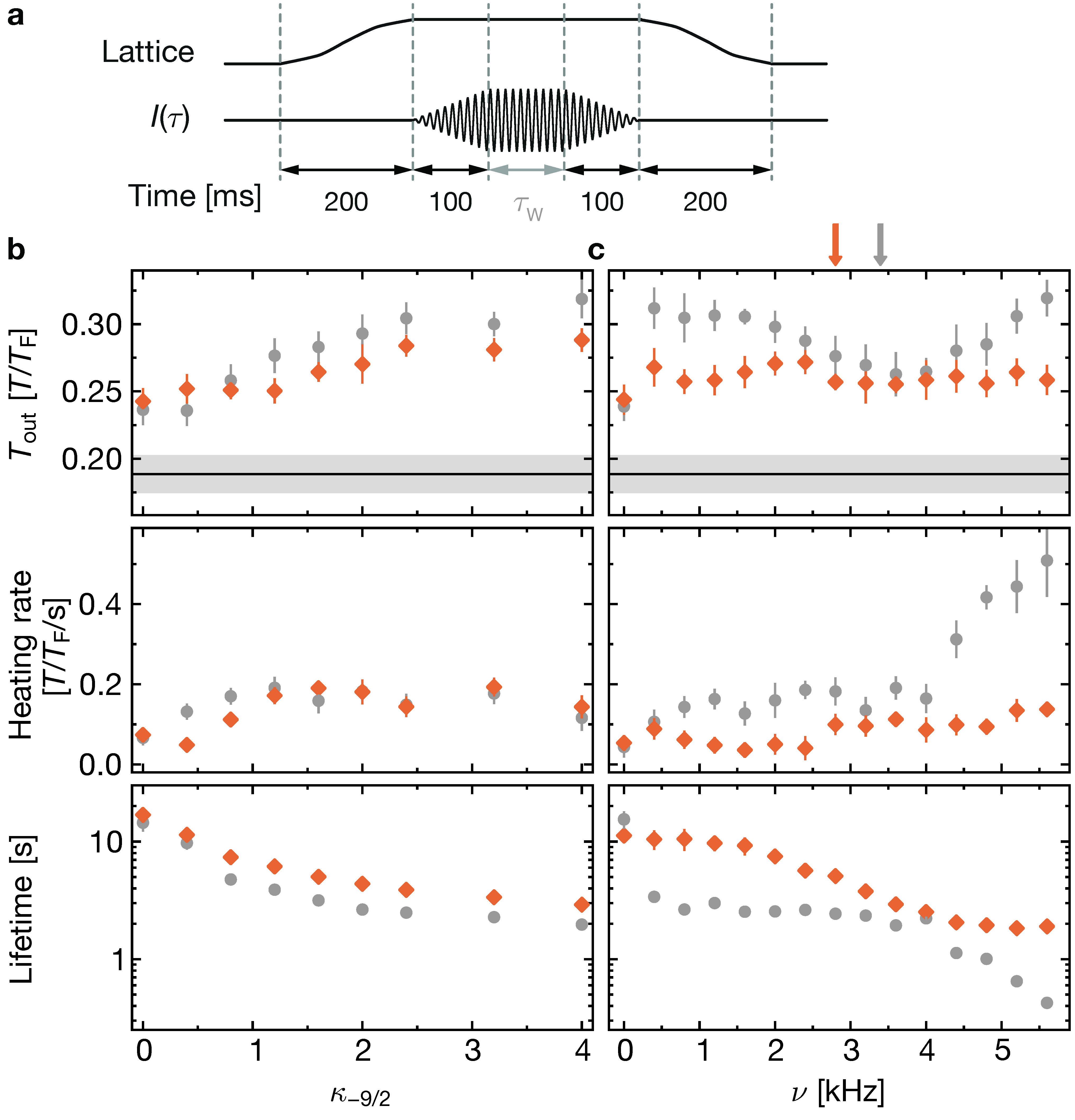}
    \caption{    
    Heating measurements. 
    {\bf (a)} 
    Schematic of the ramp protocol. 
    {\bf (b)}  
    Temperature after the full ramp ($\tau_\mathrm{W}=0$), heating rates and lifetimes as a function of modulation amplitude. 
    Grey circles (orange diamonds) indicate a lattice with tunnelling $t=h\times 174\,$Hz ($67\,$Hz). 
    The line (shaded region) show the results (error bars) in the dipole trap without the optical lattice and gradient modulation.
    {\bf (c)}  
    Same quantities measured as a function of modulation frequency, for $\kappa_{-9/2} = 1.0$. 
    Zero frequency indicates no modulation.
    Data and error bars are mean $\pm$ s.d. of 10 measurements (upper panels), or the results of linear (center) or exponential (lower) fits to the temperature and atom number \textit{versus} $\tau_{\mathrm{W}}$.
	}\label{fig:Heating}
\end{figure}

A question of fundamental interest when studying effective Floquet Hamiltonians concerns the evolution from a static to a modulated Hamiltonian as well as the stability of interacting Floquet systems\,\cite{Zenesini2009,Goldman2014,Bilitewski2015}.
From an experimental point of view, excitations created by modulating the system can be observed as heating and atom loss, both of which are detrimental.
In order to study how heating and losses depend on the amplitude and frequency of the modulation, we start with $233(10)\times10^3$ atoms in a balanced mixture of $\left| -9/2\right\rangle$ and $\left| -5/2\right\rangle$ at a temperature of $19(1)\%$ of the Fermi temperature $T_{\mathrm{F}}$. 
We turn on a three-dimensional cubic optical lattice which is well described by a single band Hubbard Hamiltonian with an isotropic tunnelling energy of either $t=h\times 174(9)\,$Hz or $67(3)\,$Hz within $200\,$ms and then linearly ramp up the modulation (which only acts in the $x$ direction) in $100\,$ms.
After a variable waiting time $\tau_{\mathrm{W}}$ this loading procedure is reverted (see Fig.\,\ref{fig:Heating}a). The final temperature after thermalization is extracted from Fermi fits to the momentum distribution\,\cite{Supplemental}.
The measurements were taken at a static field of $160.44(1)\,$G, where the scattering length of $194.8(1)a_0$ is modulated by less than $\pm 7a_0$ for the modulation parameters of Fig.\,\ref{fig:Expansion}.

Fig.\,\ref{fig:Heating}b shows how the ramp-induced heating depends on the modulation amplitude, when setting a frequency of $750\,$Hz, as above.
For small modulation amplitudes, almost no additional heating compared to the effects of the lattice-ramp itself are observed, especially for the deeper lattice.
For larger amplitudes, some heating becomes visible at this frequency, which interestingly seems to saturate when $\kappa_{-9/2} \gtrsim 2.4$ and the effective tunneling becomes small.
A heating rate is extracted from linear fits to the temperature as a function of $\tau_\mathrm{W}$ for times up to $300\,$ms (after that, the temperature starts to saturate). 
It shows negligible values for most experimentally relevant time-scales, even for strong modulation.
Exponential fits to the atom number measured for waiting times up to 2s show that modulation decreases lifetimes in the lattice.
However, even when completely localizing one species, very long lifetimes of several seconds are still observed, which correspond to values several orders of magnitude larger than the interaction and tunnelling times.

Further insight into the relevant excitations of the system can be gained by studying the dependency on modulation frequency.
The excitation of doubly occupied sites is predicted to have a resonance around $1.3\,$kHz ($2.0\,$kHz), given by the on-site interaction energy, and the width of the lowest band is $h\times 2.1\,$kHz ($0.8\,$kHz) in the shallower (deeper) lattice.
Directly above the sum of these two frequencies (marked with arrows), where such excitations should be strongly suppressed\,\cite{Greif2011}, the modulation ramp seems to cause no heating at all, see Fig.\,\ref{fig:Heating}c.
For even higher frequencies, the ramp-induced heating, heating rate and lifetime rapidly worsen for the shallower optical lattice, but not in the deeper one which has a larger bandgap.
We therefore attribute this feature to excitations of higher bands.
Although the direct gap of the non-interacting band structure has a value of $h\times 14.7\,$kHz ($21.8\,$kHz) for the shallower (deeper) lattice, an oscillating force can also drive higher-order processes.

In conclusion, we have demonstrated a versatile method for creating widely tunable state-dependent lattices with minimal heating and loss, which should be easy to implement for many existing experimental setups.
We have studied the static and dynamic behaviour of fermions in spin-dependent lattices in both real- and momentum-space.
This method makes numerous many-body Hamiltonians accessible for ultracold atoms\,\cite{Liu2004, Cazalilla2005, Zapata2010, Sotnikov2013}, including limiting cases of the spin-anisotropic Hubbard model such as the Falicov-Kimball or XXZ model.
The effects of explicitly breaking SU(2) symmetry may in fact already become visible at the level of short-range magnetic correlations observed in the Fermi-Hubbard model so far\,\cite{Greif2013, Sciolla2013, Imriska2014,Hart2015}.
In addition, interesting extensions of topologically non-trivial Hamiltonians can be accessed, such as interpolating between the Haldane and Kane-Mele models\,\cite{Haldane1988,Kane2005}.
With minor modifications our scheme could also be used to engineer gauge fields and spin-orbit coupling\,\cite{Xu2013,Anderson2013,Struck2014,Luo2015}, where circumventing spontaneous emission has been identified as a major challenge for future work\,\cite{Galitski2013}.

\begin{acknowledgments}
We thank  Ye-Hua Liu and Lei Wang for insightful discussions. We acknowledge SNF, NCCR-QSIT, QUIC (Swiss State Secretary for Education, Research and Innovation contract number 15.0019) and SQMS (ERC advanced grant) for funding. R.D. acknowledges support from ETH Zurich Postodoctoral Program and Marie Curie Actions for People COFUND program.
\end{acknowledgments}

%


\section{Supplemental Material}  

\makeatletter
\setcounter{section}{0}
\setcounter{subsection}{0}
\setcounter{figure}{0}
\setcounter{equation}{0}
\renewcommand{\thefigure}{S\@arabic\c@figure}
\renewcommand{\theequation}{S\@arabic\c@equation}
\makeatother

\subsection{Preparation of the Atomic Cloud}
We start with $1.4(4)\times10^6$ non-interacting spin-polarized fermionic $^{40}\mathrm{K}$ atoms which are cooled to temperatures of about $28(2)\%$ of the Fermi-temperature by sympathetic cooling with $^{87}\, \mathrm{Rb}$ in a magnetic QUIC-trap\,\cite{Esslinger1998}. The atoms are transferred to an optical dipole trap operating at $826\,$nm and transferred to the required spin-state using radio-frequency transitions. 
For measurements with interacting spin mixtures (Figs. 3c and 4), a mixture of $\left| -9/2\right\rangle$ and $\left| -7/2\right\rangle$ atoms is evaporated in the optical dipole trap and the $\left| -7/2\right\rangle$ atoms are then transferred to the $\left| -5/2\right\rangle$ state.
For measurements with spin-polarized non-interacting clouds (Figs. 2 and 3b), the depth of the optical dipole trap is simply lowered until the desired atom number and a narrow momentum distribution is reached.

For the measurements in Fig.\,2, the atoms are additionally levitated using a static magnetic gradient (gravity points along the $z$-direction).
In Fig.\,2, the overall harmonic confinement is $\omega_{x,y,z}=2\pi\times(15.6(1),27.8(2),54.0(5))\,$Hz and 
in Fig.\,3 it is $\omega_{x,y,z}=2\pi\times(67.8(3),60.4(4),233.5(3))\,$Hz before the expansion starts and $\omega_{x,y,z}=2\pi\times(0,58(1),124(2))\,$Hz during the expansion.
In Fig.\,4 it is $\omega_{x,y,z}=2\pi\times(55(3),52(3),112(3))\,$Hz in the lattice with $t=h\times 174(9)\,$Hz, which has a potential depth of $7\,E_{\mathrm{R}}$ per lattice beam ($E_{\mathrm{R}}=h^2/(2m_{\mathrm{K}}\lambda^2)$ denotes the recoil energy of $^{40}\mathrm{K}$ with mass $m_{\mathrm{K}}$ in a $\lambda=1064\,$nm lattice) and $\omega_{x,y,z}=2\pi\times(66(4),61(4),118(4))\,$Hz in the lattice with $t=h\times 67(3)\,$Hz, which has a depth of $11\,E_{\mathrm{R}}$.

\subsection{Gradient and Field Calibration}

All gradients are calibrated using Bloch oscillations of the $\left| -9/2\right\rangle$ and $\left| -5/2\right\rangle$ atoms in a static gradient. 
The dimensionless modulation amplitude $\kappa_{\sigma}$ is then found by dividing the Bloch-oscillation frequency by the modulation frequency.
When working at low external offset-fields, the changing offset-field arising from modulation with a single coil (which forms part of the QUIC setup) leads to a non-linear dependence of the gradient on the driving current.
A static gradient which would arise from the non-linearity is compensated.
The non-linearity also introduces higher harmonics in the modulated force. However, it does not modify the effective tunnelling by more than $2\%$.

In Fig.\,2, the levitation of the atoms leads to a different external field depending on which of the two spin-states is supported against gravity. Therefore, driving a current through the coil results in a different magnetic gradient at the position of the atoms for the two different configurations, which is taken into account in our calibration. Finally, the inhomogeneity of the gradient may lead to a spatially varying $\kappa$. For the offset field used in Fig. 3 (208.15 G), we use the damping of singlet-triplet oscillation performed as in \cite{Greif2013} to determine the variation of the gradient to be below 1 \% over the extent of the cloud.

The static and time-dependent magnetic fields are calibrated by spin-flips from the $\left| -7/2\right\rangle$ to the $\left| -5/2\right\rangle$ states as well as optical absorption measurements on the D2 line.
Note that the time-dependence of the magnetic field could be avoided by modulating the current of more than a single coil, if for example a modulation of the scattering length is not desired. The minimal change in magnetic field is then given by the applied gradient multiplied by the size of the cloud, which is typically less than $0.1\,$G.

\subsection{Quasimomentum Distribution and Dipole Oscillations}

In the experiment, the quasimomentum distribution is measured by ramping down the modulation amplitude in $10\,$ms and turning off the lattice in $0.5\,$ms, slow enough that higher bands are not populated in this ramp but fast enough that the harmonic trapping potential does not change the quasimomentum distribution during the ramp.
After $15\,$ms of ballistic expansion, an absorption image of the cloud is taken.

For Fig.\,2b, the second moment $w_q$ of the quasimomentum distribution summed over $q_y$ and $q_z$, $n(q_x)$, is determined according to $w_q^2=\frac{1}{N} \int (q_x-\left\langle q_x\right\rangle)^2 n(q_x)\mathrm{d}q_x$, where $\left\langle q_x\right\rangle=\frac{1}{N}\int q_x n(q_x)\mathrm{d}q_x$ is the mean quasimomentum and $N$ the total particle number.

The dipole oscillations in the harmonic trap are initiated by displacing the atoms by $q_0$ in quasimomentum space. As a result of the tight-binding dispersion, the equation of motion for the quasimomentum is equivalent to the one of a mathematical pendulum. The square of the oscillation frequency is proportional to the tunnelling energy $t$ and can be written as $\nu^2 = \alpha(q_0)\pi^2\nu^2_x t/E_{\mathrm{R}}$, where $\nu_x$ is the trap frequency without lattice. The parameter $\alpha(q_0)$ describes the effect of the anharmonicity of the dispersion and depends on the initial displacement. For $q_0\rightarrow 0$, the time evolution can be seen as an oscillation in a quadratic dispersion with an effective mass $m_{\mathrm{eff}}=E_{\mathrm{R}}/(\pi^2 t)\,m$ and $\alpha(q_0=0)=1$. For a finite displacement, $\alpha$ can be calculated numerically and has a value of $0.89(4)$ for our case with $q_0=0.31(4)\, \pi/a$. Therefore, we expect a frequency of $\nu_0=9.2(2)\,$Hz for the lattice configuration in Fig. 2 without gradient modulation, which is in close agreement with the experimental value of $8.4(3)\,$Hz. We attribute the residual deviations from the theoretical value to an uncertainty in the calibration of the lattice depths, which however cancels when considering $t_{\sigma}^{\mathrm{eff}}/t$. Since the displacement in quasimomentum space is the same both with and without gradient modulation, the factor $\alpha$ cancels when calculating the ratio between the two oscillation frequencies in Fig.\,2c and $(\nu/\nu_0)^2=t_{\sigma}^{\mathrm{eff}}/t$.

As the cloud has a finite width in quasimomentum space, the atoms which are closer to the edge of the Brillouin zone have a longer oscillation period since $\alpha(q_0)$ decreases as the displacement increases. While the dynamics of the peak of the distribution still follows the simple evolution described above, the motion of the center of mass depends on the exact quasimomentum distribution. In particular, the center of mass oscillates with a lower frequency than the peak. 

For the dipole oscillation data in Fig.\,2c we determine the oscillation frequency by fitting the peak of the quasimomentum distribution in the lattice, summed over $q_{y}$ and $q_{z}$.
Within this procedure we first obtain a smoothed quasimomentum distribution by applying a Savitzky-Golay filter to the raw absorption image. 
In a second step we determine the peak position of the cloud by performing a center of mass evaluation for data points which are above a threshold of $0.85$ of the maximum atomic density.
The oscillation frequency is subsequently extracted from a damped sine fit function to the peak position at different waiting times in the shaken lattice.
We obtain the error on the oscillation frequency by fitting the oscillations frequency for a threshold of $0.7$ and $1.0$ of the maximum atomic density and following the same procedure. 

\subsection{Expansion Measurements}

For the expansion data the real-space density is measured without switching off the lattice or the modulation, with the measurement time chosen such that it corresponds to a zero-crossing of the magnetic field modulation.
We measure the different spin-states separately by making use of the differential shift of the optical transition frequency used for imaging at strong external magnetic fields.
The width of the cloud $w_x$ is determined as the square root of the variance of a Gaussian fit to the \textit{in-situ} density profile.


\subsection{Heating Measurements}

We have observed that losses become very large when the magnetic offset-field (which, as mentioned above, also varies during the modulation) reaches the Feshbach resonance at $224.21(5)\,$G\,\cite{Regal2003}. 
Therefore, all heating measurements were taken at a static field of $160.44(1)\,$G, where the scattering length of $194.8(1)a_0$ is modulated by less than $\pm 7a_0$.
After the lattice has been completely ramped down, we move to an offset-field of $214.8(1)\,$G within $100\,$ms and wait for $200\,$ms in this configuration, where the scattering length is $315(2)$ Bohr radii, to ensure that any residual excitations of the cloud have thermalized.

The heating of the atoms caused by ramping into the lattice and reverting the loading procedure again gives an estimate of the actual temperature in the lattice.
In previous work it has been found that the heating caused by the first half of the ramp is lower than that of the reversed ramp\,\cite{Greif2011,Imriska2014}.

When measuring the heating for long waiting-times $\tau_{\mathrm{W}}$, we observe a saturation of $T/T_{\mathrm{F}}$.
For the data in Fig.\,4 we fit linear slopes to times up to $300\,$ms, which are the most relevant time-scales for the majority of optical lattice experiments.
Fitting to longer times would result in lower heating-rates, but the fitting function would not capture the time-dependence well.
The $1/e$-lifetimes are extracted from exponential fits to the atom numbers for waiting-times up to $2\,$s.

\end{document}